\documentstyle[prl,aps]{revtex}
\begin{document}
\draft
\title{From inflation to a zero cosmological constant phase 
without fine tuning}
\author{E. I. Guendelman\thanks{GUENDEL@BGUmail.BGU.AC.IL} and
        A. B. Kaganovich \thanks{ALEXK@BGUmail.BGU.AC.IL}}
\address{Physics Department, Ben Gurion University of the Negev,
   Beer  Sheva 84105, Israel}
\maketitle

\begin{abstract}
We show that it is possible to obtain inflation and also solve the 
cosmological constant problem. The theory is invariant under changes of 
the Lagrangian density $L$ to $L+const$. Then the constant part of a scalar 
field potential $V$ cannot be responsible for inflation. However, we show 
that inflation can be driven by a condensate of a four index field strength. 
A constraint appears which correlates this condensate to $V$. After a 
conformal transformation, the equations are the standard GR equations with 
an effective scalar field potential $V_{eff}$ which has generally an absolute 
minimum $V_{eff}=0$ independently of $V$ and without fine tuning. We also 
show that, after inflation, the usual reheating phase scenario (from 
oscillations around the absolute minimum) is possible.

\end{abstract}
\bigskip

One of the biggest puzzles of modern physics is what is referred to as the 
"cosmological constant problem", i.e. the absence of a possible constant 
part of the vacuum energy in the present day universe \cite{NWN}. On the 
other hand, many questions in modern 
cosmology appear to be solved by the so called "inflationary model" which 
makes use of a big effective cosmological constant in the 
early universe \cite{Gut}. A possible conflict between a successful 
resolution of the cosmological constant problem and the existence of 
an inflationary phase could be a "potential Achilles heel for the 
scenario" as has been pointed out\cite{KT}. Here we will see 
that indeed there is no conflict between the existence of an 
inflationary phase and the disappearance of the cosmological constant  in 
the later phases of cosmological evolution (without the need of fine 
tuning).

In Refs. \cite{GK1},\cite{GK2},\cite{GK3} we have developed an approach 
where the cosmological constant problem is treated as the absence of 
gravitational effects of a possible constant part of the Lagrangian 
density. In order to achieve this, we assume that the measure of 
integration in 
the action is not necessarily $\sqrt{-g}$ \quad ($g=Det(g_{\mu\nu})$) 
but it is determined dynamically through additional degrees of freedom. This 
theory is based on the demand that such measure respects  the principle of 
nongravitating vacuum energy (NGVE principle) \cite{GK1} which states that 
the 
Lagrangian density $L$ can be changed to $L+constant$ without affecting 
the dynamics. This requirement is imposed in order to offer a new 
approach for the solution of the cosmological constant problem.
Clearly this invariance is 
achieved if the measure of integration in the action is a total 
derivative. This is  satisfied  if 
the measure appropriate to the integration in the space of $4$ 
scalar fields  $\varphi_{a},\quad a=1,...,4$, that is 
$dV = 
d\varphi_{1}\wedge 
\ldots\wedge 
d\varphi_{4}\equiv\frac{\Phi}{4!}d^{4}x$ where
$\Phi \equiv \varepsilon_{a_{1}\ldots a_{4}}
\varepsilon^{\mu_{1}\ldots \mu_{4}}
(\partial_{\mu_{1}}\varphi_{a_{1}})
 \ldots
(\partial_{\mu_{4}}\varphi_{a_{4}})$. 

There are two well known variational principles: the
first and the second order formalisms, 
which are equivalent in the case of the general theory of relativity (GR). 
However, 
as it was shown \cite{GK2}, the first and the second order 
formalisms (starting from the same looking form of Lagrangian) are not 
equivalent in the context of the NGVE theories. The NGVE theory in
the first order formalism leads to the resolution of the cosmological 
constant problem in a straightforward way \cite{GK2},\cite{GK3} and in 
this paper, we will follow this approach. Furthemore, in the spirit of 
the theory, which assumes independence of the measure degrees of freedom 
from $g_{\mu\nu}$, the first order formalism where the independence of 
$\Gamma^{\lambda}_{\mu\nu}$ and $g_{\mu\nu}$ in the action is assumed, is 
of course much more natural.

According to the NGVE-principle, the total action in the 4-dimensional 
space-time should be written in the form $S=\int\Phi Ld^{4}x$.
 We assume that $L$ does not contain the measure fields $\varphi_{a}$, 
that is 
the fields by means of which $\Phi$ is defined. If this condition is 
satisfied then the theory has an additional symmetries \cite{GK1}. We start 
from the total Lagrangian density $L=\kappa^{-1}R(\Gamma,g)+L_{m}$,
where $L_{m}$ is the matter Lagrangian density and $R(\Gamma,g)$ is the 
scalar  curvature 
$R(\Gamma,g)=g^{\mu\nu}R_{\mu\nu}(\Gamma)$ of the space-time with the 
affine connection  $\Gamma^{\lambda}_{\mu\nu}$ (the first order 
formalism is used,that is in the action, the connection 
$\Gamma^{\lambda}_{\mu\nu}$ 
and the metric $g_{\mu\nu}$ are dynamically independent variables), \ 
$R_{\mu\nu}(\Gamma)=R^{\alpha}_{\mu\nu\alpha}(\Gamma)$, \ 
$R^{\alpha}_{\mu\nu\beta}(\Gamma)\equiv 
\Gamma^{\alpha}_{\mu\nu,\beta}+
\Gamma^{\alpha}_{\lambda\beta}\Gamma^{\lambda}_{\mu\nu}
-(\nu\leftrightarrow\beta)$.
This curvature tensor is invariant under the  
$\lambda$- transformation   \cite{E} \
$\Gamma^{\prime\alpha}_{\mu\nu}=
\Gamma^{\alpha}_{\mu\nu}+\delta^{\alpha}_{\mu}\lambda,_{\nu}$. 
 Its importance  in the 
NGVE-theory is that  allows us to eliminate the contribution to the 
torsion which appears as a result of introduction of the new measure 
\cite{GK3}.
However, even after we fix a gauge where this contribution to the torsion 
disappears, still there is the non metric contribution \cite{GK3} to the 
symmetric 
connection related to the measure. In fact, solving the equations 
following from the variation of the action with respect to 
$\Gamma^{\lambda}_{\mu\nu}$ (in the case where $L_{m}$ does not depend on 
the connection $\Gamma^{\lambda}_{\mu\nu}$) and making use the 
appropriate $\lambda$- transformation, we get 
\begin{equation}
\Gamma^{\lambda}_{\mu\nu}=
\{^{\lambda}_{\mu\nu}\} +\Sigma^{\lambda}_{\mu\nu}
\label{gam}
\end{equation}
where $\{^{\alpha}_{\mu\nu}\}$
are the 
Christoffel's connection coefficients and 
\begin{equation}
\Sigma^{\lambda}_{\mu\nu}(\sigma)=\frac{1}{2}(\delta^{\lambda}_{\mu}
\sigma,_{\nu}+
\delta^{\lambda}_{\nu}\sigma,_{\mu}-
\sigma,_{\alpha}g_{\mu\nu}g^{\lambda\alpha})
\label{sig}
\end{equation}
\begin{equation}
\sigma\equiv\ln\chi, \quad \chi\equiv\Phi/\sqrt{-g}.
\label{chi}
\end{equation}

In addition to this, in the vacuum and in some matter models, the theory 
possesses a local symmetry which plays a major role. This symmetry 
consists of a conformal transformation of the metric  
$g^{\prime}_{\mu\nu}(x)=J(x)g_{\mu\nu}(x)$ accompanied by a 
corresponding 
diffeomorphism $\varphi_{a}\rightarrow\varphi^{\prime}_{a}=
\varphi^{\prime}_{a}(\varphi_{b})$ in the space of the scalar fields
$\varphi_{a}$ such that  $J=
Det(\frac{\partial\varphi^{\prime}_{a}}{\partial\varphi_{b}})$.
Then for $\Phi$ we have: $ \Phi^{\prime}(x)=J(x)\Phi(x)$. In 
the 
presence of fermions this symmetry is appropriately generalized \cite{GK3}.
For models where it
holds it is possible to choose a gauge where the measure $\Phi$
coincides with the measure of general relativity $\sqrt{-g}$. This is
why we call this symmetry {\em "local Einstein symmetry"} (LES). 

 Varying the action with respect to $\varphi_{a}$ we get
$A^{\alpha}_{b}\partial_{\alpha}\lbrack -\frac{1}{\kappa}R(\Gamma,g)+
L_{m}\rbrack =0$ where $A^{\alpha}_{b}=\varepsilon_{a_{1}a_{2}a_{3}b}
\varepsilon^{\mu_{1}\mu_{2}\mu_{3}\alpha}
(\partial_{\mu_{1}}\varphi_{a_{1}})(\partial_{\mu_{2}}\varphi_{a_{2}}) 
(\partial_{\mu_{3}}\varphi_{a_{3}})$. Since 
$A_{b}^{\alpha}\partial_{\alpha}\varphi_{b^{\prime}}=\frac{1}{4}\delta_{bb^{\prime}}\Phi$
it follows that  $Det (A_{b}^{\alpha}) = 
\frac{4^{-4}}{4!}\Phi^{3}$. If $\Phi\neq 0$, we obtain
\begin{equation}
-\frac{1}{\kappa}R(\Gamma,g)+L_{m}=M=constant.
\label{1Int}
\end{equation}

Varying the action with respect to 
$g^{\mu\nu}$ we get (for simplicity we present here the calculations for the 
case
when there are no fermions)
 \begin{equation}
-\frac{1}{\kappa}R_{\mu\nu}(\Gamma)+
\frac{\partial L_{m}}{\partial g^{\mu\nu}}=0.
\label{RAB}
\end{equation}

In the case where $L_{m}$ does not depend on $\Gamma^{\alpha}_{\mu\nu}$, 
Eq. (\ref{1Int}) takes the form 
\begin{equation}
\Box\chi^{1/2}-\frac{1}{6}[R(g)-\kappa (L_{m}-M)]\chi^{1/2}=0
 \label{SL}
 \end{equation}
where $R(g)$ is the Riemannian scalar curvature.

Contracting Eq. (\ref{RAB}) with $g^{\mu\nu}$ 
and making use Eq. (\ref{1Int}) we get the constraint
\begin{equation}
g^{\mu\nu}\frac{\partial(L_{m}-M)}{\partial g^{\mu\nu}}-(L_{m}-M)=0.
\label{con}
\end{equation}

For the cases where the LES is an exact symmetry, we 
can eliminate the mentioned above $\chi$-contribution to the connection. 
Indeed, for  $J=\chi$ we get $\chi^{\prime}\equiv 1$ and 
$\Gamma^{\prime \alpha}_{\mu\nu}=
\{ ^{\alpha}_{\mu\nu}\}^{\prime}$, where
$\{ ^{\alpha}_{\mu\nu}\}^{\prime}$
 are the Christoffel's coefficients corresponding
to the new metric $g^{\prime}_{\mu\nu}$. In this gauge the metric-affine 
space-time becomes a Riemannian space-time (in the absence of fermions).

In the presence of matter, the LES may be lost. However, the theory still 
makes sense \cite{GK2} and the resolution of the cosmological constant 
problem is 
retained. Together with this, the LES appears to have a deep geometric 
meaning \cite{GK3}. 

Now let us consider cases when the constraint (\ref{con}) is not satisfied
without restrictions on the dynamics of the matter fields. Nevertheless,
the constraint (\ref{con}) holds as a consequence of the variational
principle in any situation.

A simple situation where the constraint (\ref{con}) is not automatic is the 
case
of a single scalar field with a nontrivial potential $V(\phi)$. In this 
case the constraint (\ref{con}) implies
\begin{equation}
V(\phi)+M=0.
 \label{CS}
 \end{equation}
Therefore we conclude that, provided $\Phi\neq 0$, there is no dynamics for
the theory of a single scalar field, since the constraint (\ref{CS}) forces
this scalar field to be a constant \cite{GK2}. Inserting (\ref{CS}) into 	
(\ref{SL}), we obtain $\Box\chi^{1/2}-\frac{1}{6}R(g)\chi^{1/2}=0$. This 
conformal coupling of $\chi^{1/2}$ with $R(g)$ shows that on the mass 
shell the LES is restored (this can be seen in all the other equations as 
well). Making use then of the gauge $\chi =1$, we 
see that $R(g)=0$ and therefore the only maximally symmetric solution is 
Minkowski space. Consistency with the scalar field equation requires 
$V^{\prime}\equiv\frac{\partial V}{\partial\phi}=0$, 
that is $\phi$ is located on an extremum of $V(\phi)$.

As it follows from our analysis above, a model with only a scalar field  
cannot give rise to inflation since the gravitational effects of the scalar 
field potential is always cancelled by the integration constant $M$. We will 
see  that  nontrivial dynamics of a single scalar field including the 
possibility of inflation  can be 
obtained by considering a model with an additional degree of freedom 
described by a three-index potential $A_{\beta\mu\nu}$  as in the Lagrangian 
density 
\begin{equation}
L=-\frac{1}{\kappa}R(\Gamma,g) 
+\frac{1}{2}\phi,_{\alpha}\phi^{,\alpha}-V(\phi)+
\frac{1}{4!}F_{\alpha\beta\mu\nu}F^{\alpha\beta\mu\nu}. 
 \label{L}
 \end{equation}  
Here $F_{\alpha\beta\mu\nu}\equiv\partial_{[\alpha}A_{\beta\mu\nu ]}$ is 
the field strength which is invariant under the gauge transformation 
$A_{\beta\mu\nu}\rightarrow A_{\beta\mu\nu}+\partial_{[\beta}f_{\mu\nu]}$. 
In ordinary 4-dimensional GR, the 
$F_{\alpha\beta\mu\nu}F^{\alpha\beta\mu\nu}$ term gives rise 
to a  cosmological constant depending on an integration constant 
\cite{FF}. In our case, 
due to the constraint (\ref{con}), the degrees of freedom contained in 
$F_{\alpha\beta\mu\nu}$ and those of the scalar field $\phi$ will be
intimately correlated. The sign in front of the 
$F_{\alpha\beta\mu\nu}F^{\alpha\beta\mu\nu}$ term is chosen so that in 
this model the 
resulting expression for the energy density of the scalar field $\phi$ is 
a positive definite one for any possible 
space-time dependence of $\phi$ in an effective "Einstein picture" (see 
below). Notice also 
that two last terms in the action with the Lagrangian (\ref{L}) break 
explicitly the LES.

The gravitational equations (\ref{RAB}) take now the form   
\begin{equation}
-\frac{1}{\kappa}R_{\mu\nu}(\Gamma)
+\frac{1}{2}\phi,_{\mu}\phi,_{\nu}+
\frac{1}{6}F_{\mu\alpha\beta\gamma}F_{\nu}^{\alpha\beta\gamma}=0.
 \label{RF}
 \end{equation}
Notice that the scalar field potential $V(\phi)$ does not appear 
explicitly in Eqs. (\ref{RF}). However, the constraint (\ref{con}), which 
takes now  the form
\begin{equation}
V(\phi)+M=
-\frac{1}{8}F_{\alpha\beta\mu\nu}F^{\alpha\beta\mu\nu},
 \label{conF}
 \end{equation}
allows us to express the last term in (\ref{RF}) in terms of the 
potential $V(\phi)$ (using the fact that 
$F^{\alpha\beta\mu\nu}\propto\varepsilon^{\alpha\beta\mu\nu}$ in 
4-dimensional space-time).

Varying the action with respect to $A_{\nu\alpha\beta}$, we get the equation
$\partial_{\mu}(\Phi F^{\mu\nu\alpha\beta})=0$. Its general solution has 
a form (see definition (\ref{chi})).
\begin{equation}
F^{\alpha\beta\mu\nu}=
\frac{\lambda}{\Phi}\varepsilon^{\alpha\beta\mu\nu}
\equiv\frac{\lambda}{\chi\sqrt{-g}}\varepsilon^{\alpha\beta\mu\nu} ,
 \label{F}
 \end{equation}
where $\lambda$ is an integration constant. Then 
$F_{\alpha\beta\mu\nu}F^{\alpha\beta\mu\nu}=-\lambda^{2}4!/\chi^{2}$ and 
therefore
\begin{equation}
V(\phi)+M=3\lambda^{2}/\chi^{2}
 \label{V+M}
 \end{equation} 
and 
$F_{\mu\alpha\beta\gamma}F_{\nu}^{\alpha\beta\gamma}=
-(6\lambda^{2}/\chi^{2})g_{\mu\nu}=-2[V(\phi)+M]g_{\mu\nu}$ (notice that 
$F_{\alpha\beta\mu\nu}F^{\alpha\beta\mu\nu}$ is not a constant now as 
opposed to the GR case \cite{FF}). This shows how 
the potential $V(\phi)$ appears in Eq. (\ref{RF}), spontaneously violating 
the symmetry of the action $V(\phi)\rightarrow V(\phi)+constant$, which 
now corresponds to a redefinition of the integration constant $M$.

The equation of motion of the scalar field $\phi$ is 
$(-g)^{-1/2}\partial_{\mu}(\sqrt{-g}g^{\mu\nu}\partial_{\nu}\phi)
+\sigma,_{\mu}\phi ^{,\mu}+V^{\prime}(\phi)=0$, where 
$V^{\prime}\equiv\frac{\partial V}{\partial\phi}$.

The derivatives of the field 
$\sigma$ enter both in the gravitational Eqs. (\ref{RF}) 
(through the connection) and in the scalar field equation. In order to 
see easily the physical content of this model, we have to perform a 
conformal transformation $\overline{g}_{\mu\nu}(x)=\chi 
g_{\mu\nu}(x)$ to obtain an "Einstein picture". 
Notice that now this transformation is not a symmetry and indeed 
changes the form of equations.  In 
this new frame, the gravitational equations become those of GR in the 
Riemannian space-time with metric $\overline{g}_{\mu\nu}$

\begin{equation}
R_{\mu\nu}(\overline{g}_{\alpha\beta})-
\frac{1}{2}\overline{g}_{\mu\nu}R(\overline{g}_{\alpha\beta})=
\frac{\kappa}{2}T^{eff}_{\mu\nu}(\phi),
 \label{GR}
 \end{equation}
where 
the source is 
the minimally coupled scalar field $\phi$
\begin{equation}
T^{eff}_{\mu\nu}(\phi)=\phi_{,\mu}\phi_{,\nu}-
\frac{1}{2}\overline{g}_{\mu\nu}\phi_{\alpha}\phi_{\beta}
\overline{g}^{\alpha\beta}+\overline{g}_{\mu\nu}V_{eff}(\phi)
 \label{TGR}
 \end{equation}
 with the new effective potential
\begin{equation}
V_{eff}\equiv\frac{2}{\lambda 3\sqrt{3}}(V+M)^{3/2}.
 \label{V}
 \end{equation}
The scalar field equation takes a conventional form 
$(-\overline{g})^{-1/2}\partial_{\mu}(\sqrt{-\overline{g}}\enspace 
\overline{g}^{\mu\nu} \partial_{\nu}\phi)+V^{\prime}_{eff}(\phi)=0$.
Notice again that the potential $V_{eff}(\phi)$ is non negative one which 
is a result of the choice  of sign in front of the 
$F_{\alpha\beta\mu\nu}F^{\alpha\beta\mu\nu}$ term in (\ref{L}). 

We see that in the Einstein picture, for any analytic $V(\phi)$, \ 
$V_{eff}(\phi)$ has an 
extremum, that is $V^{\prime}_{eff}=0$, {\em either} when $V^{\prime}=0$ 
{\em or\/} $V+M=0$. The extremum when $V+M=0$ corresponds to an absolute 
minimum (since $V_{eff}(\phi)$ is non negative) and therefore it is {\em a 
vacuum with 
zero effective cosmological constant\/}. It should be emphasized that all 
what is required is that $V+M$ touches zero at {\em some\/} point 
$\phi_{0}$ but $V^{\prime}$ at this point does not  need to vanish. 
Therefore {\em no fine tuning\/} in the usual sense, of adjusting a 
minimum of  a potential to coincide with the point where this 
potential itself vanishes, is required. And the situation is even more 
favorable since even if $V+M$ happens not to touch  zero for any value of 
$\phi$, we  always have an infinite set of other integration constants
where this will happen.  

In the context of the cosmology, for the Friedmann- 
Robertson-Walker universe where in the Einstein picture $\phi 
=\phi(t)$ and $d\overline{s}^{2}=\overline{g}_{\mu\nu}dx^{\mu}dx^{\nu}= 
dt^{2}-\overline{a}^{2}(t)dl^{2}$, 
$dl^{2}=[dr^{2}/(1-kr^{2})+r^{2}d\Omega^{2}]$, we notice that due 
to the positivity  of $V_{eff}$, all the known inflationary scenarios 
\cite{Gut} for the very early universe can be implemented 
depending on the choice of the potential $V(\phi)$. It is very 
interesting that the parameters ruling the inflation are controlled by 
the integration constants $M$ and $\lambda$.

After inflation, when the scalar field $\phi$ approaches the position 
$\phi_{0}$ of the absolute minimum of the potential $V_{eff}$, the 
$\chi$-field approaches infinity as it seen from the constraint (\ref{V+M}).
To clarify the meaning of this effect, let us go back to the picture with 
the original $g_{\mu\nu}$ while still using the cosmic time $t$ that was 
defined in the Einstein picture. Then equation for $\phi$ is
\begin{equation}
\ddot \phi +3\frac{\dot a}{a}\dot \phi-
\frac{3V^{\prime}}{4(V+M)}\dot \phi^{2}+
\frac{\sqrt{V+M}}{\lambda\sqrt{3}}V^{\prime}=0,
 \label{FA}
 \end{equation}
where $a^{2}(t)=\overline{a}^{2}(t)/\chi(t)$, $g_{00}(t)=1/\chi (t)$ and 
constraint (\ref{V+M}) have been used.

Generally, $\dot{\phi}$ does not go to zero as $\phi\rightarrow\phi_{0}$ 
(and
$V(\phi)+M\rightarrow 0$). 
In this asymptotical region we can find the first integral of Eq. (\ref{FA}).
 Assuming 
that $V^{\prime}(\phi_{0})\neq 0$, i.e. without fine tuning, we get
\begin{equation}
\dot{\phi}a^{3}(t)\simeq c[V(\phi)+M]^{3/4},\quad c=const \quad (as\quad  
\phi\rightarrow\phi_{0}),
 \label{aVM}
 \end{equation}
which means that $a(t)\rightarrow 0$ as $\phi\rightarrow\phi_{0}$ 
(notice that if we would have  chosen a coordinate frame in
the original picture such that
$ds^{2}=
dt^{\prime 2}-a^{2}(t^{\prime})dl^{2}$, then instead of (\ref{aVM}) we 
would have gotten $a^{3}(t^{\prime})d\phi/dt^{\prime}\simeq 
c[V(\phi)+M]^{1/2}$ as
$\phi\rightarrow\phi_{0}$). Then integrating the gravitational 
equations we get asymptotically (as $\phi\rightarrow\phi_{0}$) 
that $a^{2}(t)=a_{0}^{2}/\chi(t)$, \quad $a_{0}=const$, that is in the 
original frame 
there is a collapse of the universe from a finite $a$ to $a=0$ in a 
finite time and therefore the Riemannian curvature goes to infinity  as 
$\phi\rightarrow\phi_{0}$. 
This pathology {\em is not seen} in the 
Einstein frame due to the singularity of the conformal transformation 
$\overline{a}^{2}=\chi a^{2}$ at $\phi =\phi_{0}$. In fact, this is not a 
problem from the point of view of physics, since as 
$\phi\rightarrow\phi_{0}$  (and $V(\phi)+M\rightarrow 0$), the LES 
becomes restored at the critical point $\phi\equiv\phi_{0}$ where 
$V(\phi_{0})+M=0$. In the presence of the LES, the conformal 
transformation  $\overline{g}_{\mu\nu}(x)=\chi g_{\mu\nu}(x)$
becomes part of the LES transformation and represents a nonsingular gauge 
choice.

A real problem in the scenario discussed above is the fact that at the 
point $\phi=\phi_{0}$ we have $V_{eff}=0$, $V^{\prime}_{eff}=0$ but 
$V^{\prime\prime}_{eff}$ diverges at $\phi=\phi_{0}$. This causes 
problems both in the cosmological picture when considering the possibility 
of small oscillations arround the minimum and in the associated particle 
physics, since masses of scalars, like for example the Higgs mass, will 
appear always infinite.

It turns out that the model with the Lagrangian density (\ref{L}) 
described above is only a representative of a family of possible models 
with actions $S=\int\Phi Ld^{4}x$ where 
\begin{equation}
L=-\frac{1}{\kappa}R(\Gamma,g)
+\frac{1}{2}\phi,_{\alpha}\phi^{,\alpha}-V(\phi)-
\frac{1}{4p-1}(-F_{\alpha\beta\mu\nu}F^{\alpha\beta\mu\nu})^{p},
 \label{Lp}
 \end{equation}
where still $F_{\alpha\beta\mu\nu}\equiv\partial_{[\alpha}A_{\beta\mu\nu ]}$.
Here $p\not=1/4$ is a real number parametrizing the family of Lagrangians 
(\ref{Lp}). In this case, once again solving the equation of motion 
obtained from variation with respect to $A_{\mu\nu\alpha}$ and then using 
the associated constraint that replaces (\ref{conF}) we obtain instead of 
Eq.(\ref{V+M}) 
\begin{equation}
\lambda^{2}/\chi^{2}=\frac{1}{24}(V(\phi)+M)^{2-1/p}
 \label{V+Mp}
 \end{equation}

Then instead of Eq.(\ref{V}) the associated effective potential in the 
Einstein picture is
\begin{equation}
V_{eff}\equiv\frac{1}{\lambda\sqrt{24}}(V+M)^{2-1/2p}.
 \label{Vp}
 \end{equation}

As before, the extremum when $V+M=0$ corresponds to an absolute
minimum (for any $p>1/2$) and therefore it is {\em a
vacuum with
zero effective cosmological constant\/}.
We can now notice that  the limit $p\rightarrow\infty$ is critical one, 
since in this limit $V_{eff}=\frac{1}{\lambda\sqrt{24}}(V+M)^{2}$ and for 
any analytical function $V(\phi)$, all derivatives of the 
effective potential $V_{eff}(\phi)$ are finite at the 
absolute minimum $\phi=\phi_{0}$ where $V(\phi_{0})+M=0$. In 
particular, $V^{\prime\prime}_{eff}(\phi_{0})\propto 
[V^{\prime}(\phi_{0})]^{2}$ is finite (and nonzero if we do not carry out 
the fine tuning $V^{\prime}(\phi_{0})=0$). Therefore the Higgs boson, in 
particular, can reappear  
as a physical particle of the theory. In the context of  cosmology 
where $V_{eff}(\phi)$ plays the role of the inflaton potential, a finite 
mass of the inflaton allows to recover the usual oscillatory regime of 
the reheating period after inflation that are usually considered.

The incorporation of gauge fields and fermions into the family of 
Lagrangians (\ref{Lp}) has also many interesting features in the Einstein 
picture like the appearence of normal Maxwell dynamics in the low energy 
limit in the abelian case and standard Yang-Mills behavior in the 
nonabelian case, the appearence of mass for fermions, etc.. In the 
associated particle physics, the $p\rightarrow\infty$ limit has 
remarkable features as well. These subjects will be studied in a longer 
publication \cite{GK}.

As one of the referees of this paper pointed out, the picture presented 
here should be regarded as a preliminary one. In particular, questions
that concern reheating and density perturbations have to be analyzed. 
Here, in addition to the usual possibility of choosing 
the potential of the scalar field which governs the cosmological 
processes mentioned before, this model has the additional integration 
constants $M$ and $\lambda$ which also enter in the effective 
potential. Therefore we hope that this model will provide more 
possibilities to obtain naturally the correct density perturbations 
and reheating.


\begin{thebibliography}{99}

\bibitem{NWN}
 S. Weinberg, Rev. Mod. Phys. {\bf 61}, 1
(1989); Y. J. Ng, Int. J. Mod.
Phys. {\bf D1}, 145, (1992); {\it Gravitation and Modern Cosmology, The
Cosmological Constant Problem}, edited by A. Zichichi, V. de Sabbata and N.
 Sanchez (Ettore Majorana International Science Series, Plenum Press, 1991).
For the history of the problem see for example discussion in  I. Novikov,
{\it Evolution of the Universe,} Cambridge University Press, 1983.


\bibitem{Gut}
A. H. Guth, Phys. Rev. {\bf D23}, 347 (1981); A. D. Linde, Phys. Lett. 
{\bf 108B}, 389 (1982); A. Albrecht and P. J. Steinhardt, Phys. Rev. 
Lett. {\bf 48}, 1220 (1982), A. A. Starobinsky, Phys. Lett. {\bf B91}, 99 
(1980); A. D. Linde, Phys. Lett. {\bf B129}, 177 (1983); D. La and D. J. 
Steinhardt, Phys. Rev. Lett. {\bf 62}, 376 (1989).   


\bibitem{KT}
See, for example: E. Kolb and M. S. Turner,{\it The Early Universe}, Addison 
Wesley, 1990 (see p.314).

\bibitem{GK1}
E.I. Guendelman and A.B. Kaganovich, Phys. Rev. {\bf D53}, 7020
(1996). 

\bibitem{GK2}
E.I. Guendelman and A.B. Kaganovich,  Phys. Rev.{\bf D55}, 5970 (1997).

\bibitem{GK3}
E.I. Guendelman and A.B. Kaganovich, Phys. Rev.{\bf D56} (1997).

\bibitem{E}
A.Einstein, {\it The Meaning of Relativity}, Fifth Edition, MJF
Books, N.Y.1956   ( see Appendix II).


\bibitem{FF}
A. Aurilia, H. Nicolai and P. K. Townsend, Nucl.Phys. 
{\bf B176}, 509 (1980); S. W. Hawking, Phys. Lett., {\bf 134B}, 276 
(1984);  A. Aurilia, G. Denardo, F. Legovini and E. Spalluci, ibid., {\bf 
147B}, 258 (1984); E. Witten , in Shelter II 1985 {\it Proc. 1983 Shelter 
Island 
Conference on Quantum Field Theory and the Fundamental Problems of 
Physics} (Cambridge, MA: MIT Press); J. D. Brown and C. Teitelboim, 
Phys. Lett. {\bf 195B}, 177 (1987); Nucl. Phys. {\bf B297}, 787 (1988).

\bibitem{GK}
E.I. Guendelman and A.B. Kaganovich, in preparation. 

\end{thebibliography}
\end{document}